\documentclass{styles/svproc}
\usepackage{graphicx}
\usepackage{float}
\usepackage{url}

\begin{document}
\mainmatter              

\title{Effects of Nozzle Roughness on the Streamwise Streaks in Underexpanded Jets – An Experimental Study}
\titlerunning{ISSW 35}  

\author{H. Gong, S. Wang*}

\institute{SKLTCS, CAPT, School of Mechanics and Engineering Science, Peking University, \\5 Yiheyuan Road, Haidian District, Beijing, 100871, China\\
\email{sk.wang@pku.edu.cn}}

\maketitle 

\begin{abstract}
The present study investigated the formation of streamwise streaks in underexpanded jets from round sonic nozzles, based on direct experimental observation using high-speed schlieren imaging and PLIF methods. The effect of geometric perturbations of the nozzle exit on the streamwise flow structure was examined through a series of comparative experiments. Underexpanded jets were generated in a vacuum chamber using $\phi$-4 mm nozzles of two different configurations: (a) a "smooth" nozzle, shaped and polished by a high-quality commercial lathe machine; and (b) nozzles with artificially introduced sinusoidal perturbation of various wavenumbers on the circular contour of the exit. In the case of the "smooth" nozzle, experiments were repeated following a 60-degree rotation of the nozzle along its axis, and a similar rotation in the streak patterns was observed. This suggests that the streamwise streaks most likely originated from geometric perturbations caused by the minute roughness at the nozzle exit. In the latter case, the effects of modal distribution of geometric perturbation on the streaks were further investigated. The results showed that the low-wavenumber (k $<$ 5) perturbations exhibited much smaller growth rates of streamwise streaks -- likely dominated by residual roughness similar to the "smooth" case -- compared to higher-wavenumber (k = 6 and 7) perturbations, where the streak patterns were observed to correlate geometrically with the perturbed nozzle exit contour. Results from the present study should prove useful in enhancing the current understanding of noise patterns in supersonic wind tunnel tests, where nozzles are critical components.

\end{abstract}

\section{Introduction}
The presence of streamwise streaks along the primary Mach cell structures in supersonic expanding jets was observed almost half a century ago [Glotov and Moroz, 1977], and was later found to be critical to the mixing and turbulent transition of flow near the jet boundary. Despite the prominence of such streamwise structures, a consensus on their origin has not yet been reached. Over the past 50 years, more than 30 experimental studies have investigated the streamwise structures of underexpanded jets [Franquet et al., 2015], revealing two possible causes: the intrinsic instability of supersonic flow over the concave hydrodynamic boundary of an underexpanded jet, and geometric perturbations from rough elements at the nozzle exit. The results have been somewhat contradictory. For example, Krothapalli et al. [Krothapalli et al., 1991] observed streamwise vortex structures that were not affected by disturbances originating upstream of the nozzle exit; this was corroborated by Arnette et al. [Arnette et al., 1993], who reported that the characteristics of the streamwise structure closely aligned with the theoretical properties of Görtler vortices, leading to speculation that Taylor-Görtler instability was the primary cause. More recently, however, Zapryagaev et al. [Zapryagaev et al., 2004] demonstrated that artificially introducing roughness elements at the nozzle exit can generate similar streamwise structures in the jet. To date, there has been no definitive conclusion as to whether hydrodynamic instability or nozzle roughness is the dominant cause of streamwise streaks, motivating the current study to further investigate this problem.

\section{Experimental Setup}
In the current study, underexpanded jets were generated using a set of similar nozzles with an average radius of 4 mm, inside a vacuum-pumped cylindrical flow chamber with a diameter of 130 mm and a height of 110 mm. The nozzles were supplied with gas from an upstream mixing tank with a volume of 25 L, where both temperature and pressure could be controlled. The mixing tank and flow chamber were separated by a slide gate valve. For the Schlieren experiments, pure nitrogen gas at room temperature and atmospheric pressure was used, while for the PLIF experiments, pure acetone vapor at 100°C and half an atmosphere of pressure was employed. The downstream end of the flow chamber was connected to a secondary vacuum chamber with a volume greater than one cubic meter, which helped maintain a relatively stable pressure within the flow chamber. This secondary vacuum chamber was further connected to a two-stage vacuum pump capable of achieving a maximum pumping rate of 4 L/s. The vacuum pump was operated continuously during each jet experiment, and the underexpanded supersonic jets were produced by rapidly opening the slide gate valve.

Throughout each experiment, gas gradually accumulated in the flow chamber, causing the background pressure to increase over time and resulting in unsteady flow dynamics. The corresponding flow structures were observed using side-view high-speed schlieren imaging, as well as PLIF imaging across selected planes oriented at 45 degrees to the flow direction. A schematic of the experimental setup is shown in Fig. 1. During each experiment, the chamber’s back pressure increased from less than 1 Pa to atmospheric pressure within a fraction of a second. The instantaneous pressure history was estimated from the location of the primary Mach disk, whose distance from the nozzle exit scales with the square root of the pressure ratio.

\begin{figure}[h]
\begin{center}
\includegraphics[width=\linewidth]{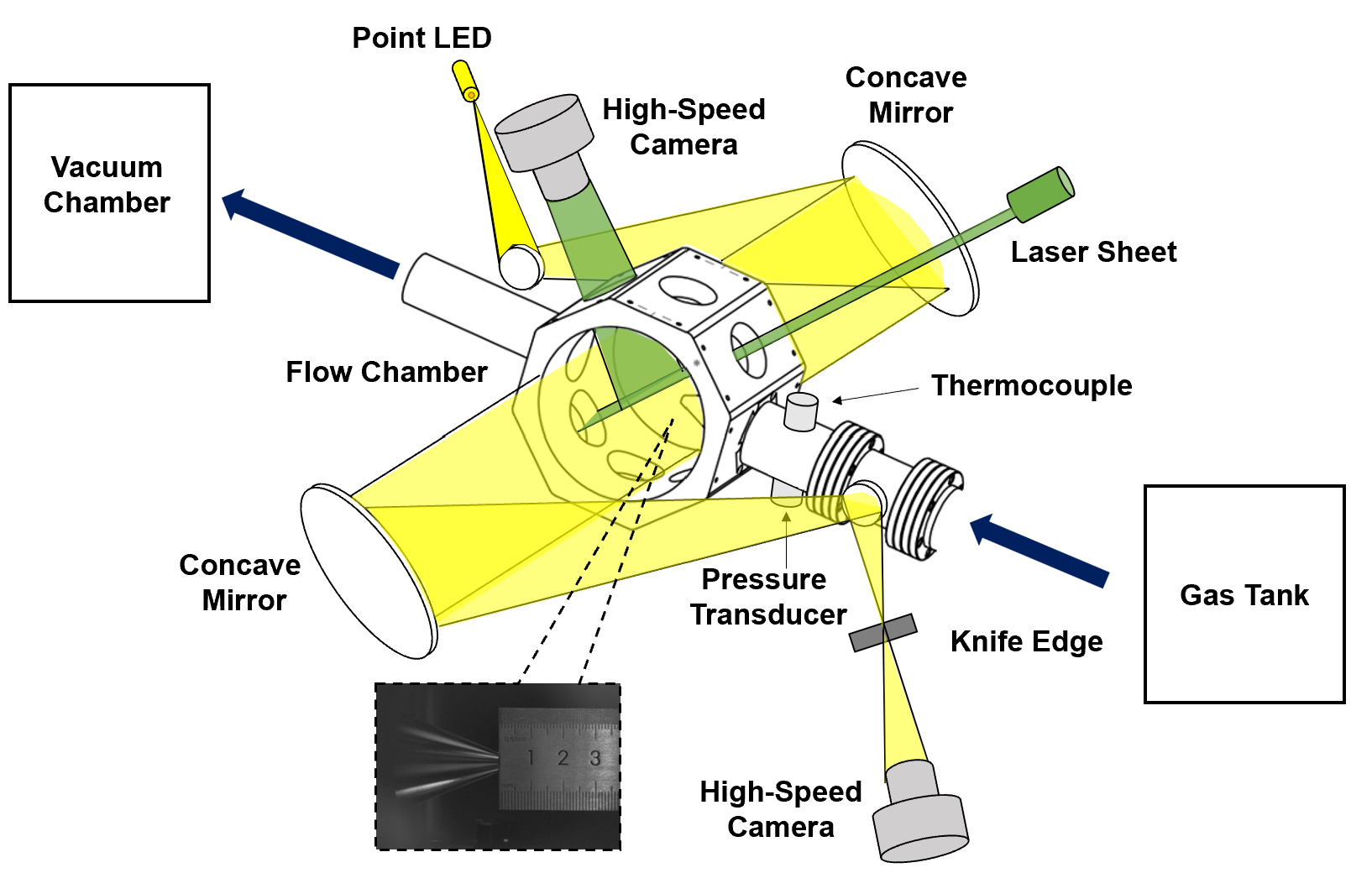}
\end{center}
\caption{\textbf{} The current experimental setup.}
\label{fig:1}
\end{figure}

Based on the smooth nozzle, the nozzle exit was modified using axisymmetric stamping knives with different wavenumbers to create nozzles featuring geometric disturbances of the corresponding modes. These modifications were intended to excite streamwise streak structures in the jet with the respective wavenumbers, allowing observation of the streamwise evolution of different streak modes under aerodynamic conditions. The wavenumbers applied in this study ranged from 3 to 7. The smooth nozzle, the nozzles with disturbances, and the corresponding stamping knives are shown in Fig. 2.
 
\begin{figure}[H]
\begin{center}
\includegraphics[width=0.8\linewidth]{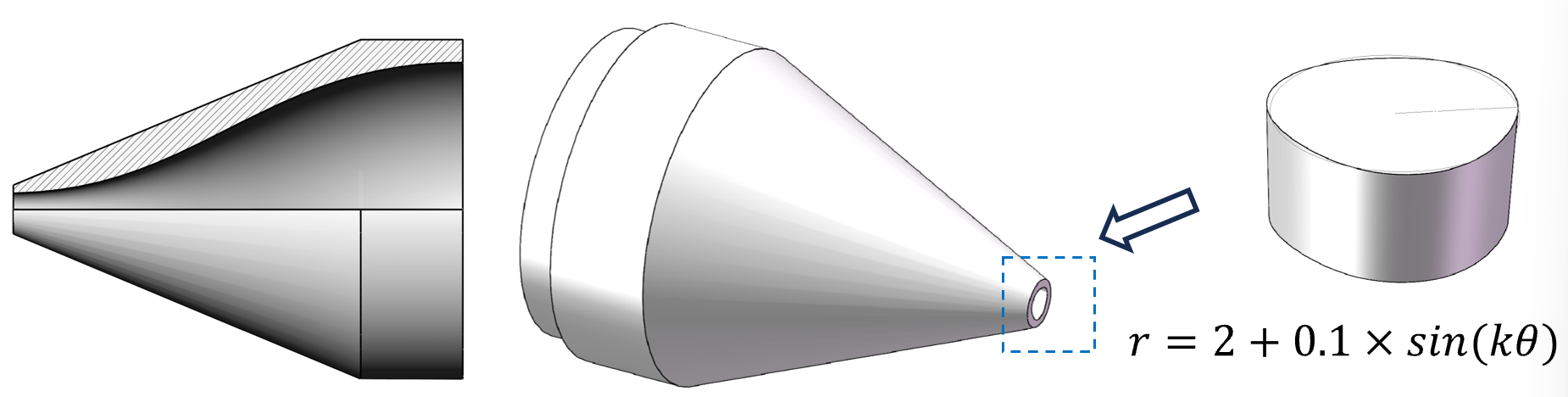}
\end{center}
\caption{\textbf{} Sonic nozzles and the stamping knives used in the current study.}
\label{fig:2}
\end{figure}

\section{Results}

Fig. 3 shows images from three repeated experiments of jets generated by the smooth nozzle. To better visualize the position, morphology, and strength of the streamwise structures, the image contrast was enhanced through digital image processing. A distinctive streak pattern was observed near the boundary of the barrel shock. This pattern was found to be relatively stable and consistent across repeated experiments under different back pressure ratios, indicating that it was probably not, or at least not entirely, induced by intrinsic hydrodynamic instabilities arising from random fluctuations in the flow field. A close examination of the "smooth" nozzle revealed the presence of micron-sized bumps and grooves on the inner surface near the exit, caused by minute machining defects. In a sense, the streamwise streak patterns of underexpanded jets were essentially "fingerprints" of the nozzle roughness, amplified by the flow.

\begin{figure}[H]
\begin{center}
\includegraphics[width=0.6\linewidth]{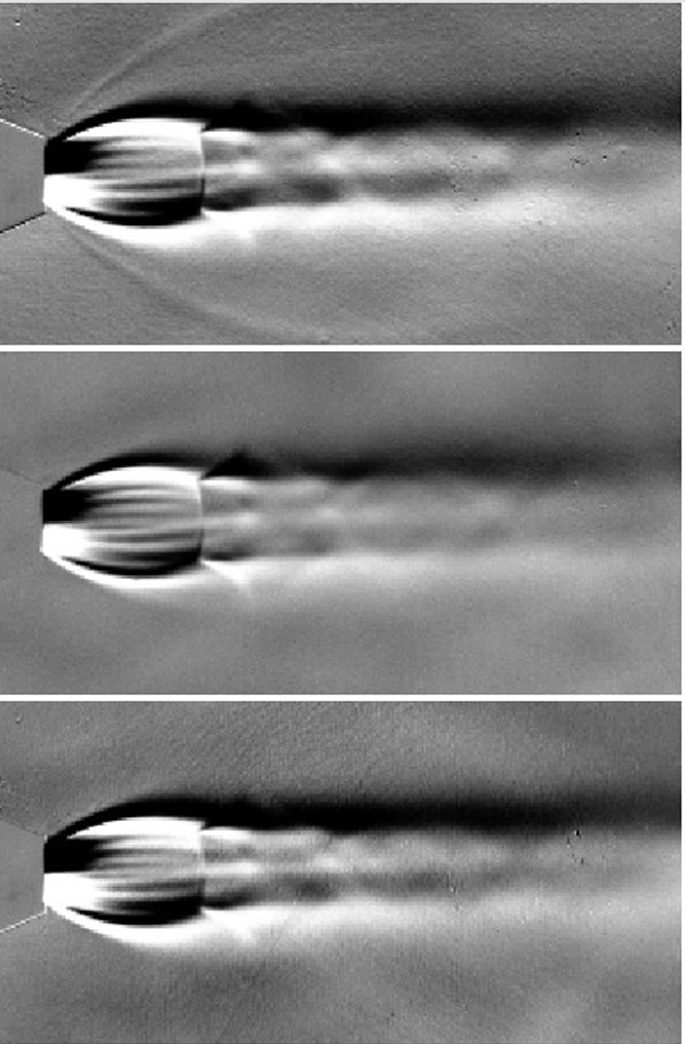}
\end{center}
\caption{\textbf{} Representative schlieren images
of three repeated experiments with a smooth nozzle.  All schlieren images were taken under conditions of L/D$_0$ = 2.5.}
\label{fig:3}
\end{figure}

An additional set of experiments was conducted using the smooth nozzle, in which the nozzle was rotated azimuthally in increments of 60 degrees while maintaining the same viewing angle. Some representative results are shown in Fig. 4. A consistent shift in the streak patterns was observed following each nozzle rotation, further corroborating that the streamwise streaks most likely originated from a fixed disturbance pattern in the nozzle geometry.

\begin{figure}[h]
\begin{center}
\includegraphics[width=\linewidth]{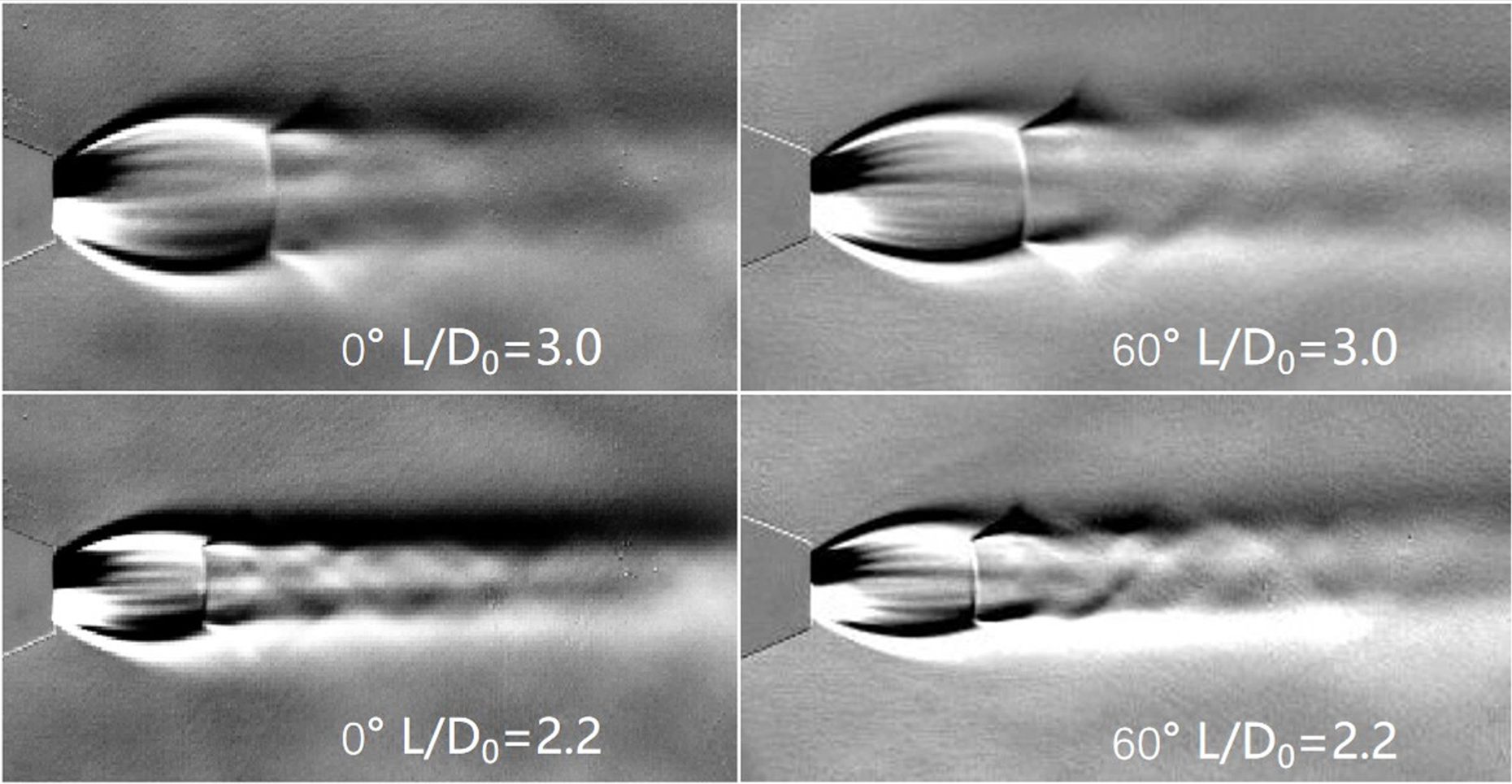}
\end{center}
\caption{\textbf{}Comparison between streamwise streak patterns of the "smooth" nozzle at different nozzle azimuth angle. The top and bottom panels correspond to different pressure ratios.}
\label{fig:4}
\end{figure}

However, it is worth noting that, although intrinsic hydrodynamic instabilities may not be the direct cause of streamwise streaks, they could still influence the streamwise development of these streaks, such as their growth rate. Nozzles with geometric disturbances of specific wavenumbers were used to excite the corresponding streak modes. Two sets of representative schlieren images of jets activated by different wavenumbers under identical conditions are shown in Fig. 5 and Fig. 6.

\begin{figure}[h]
\begin{center}
\includegraphics[width=\linewidth]{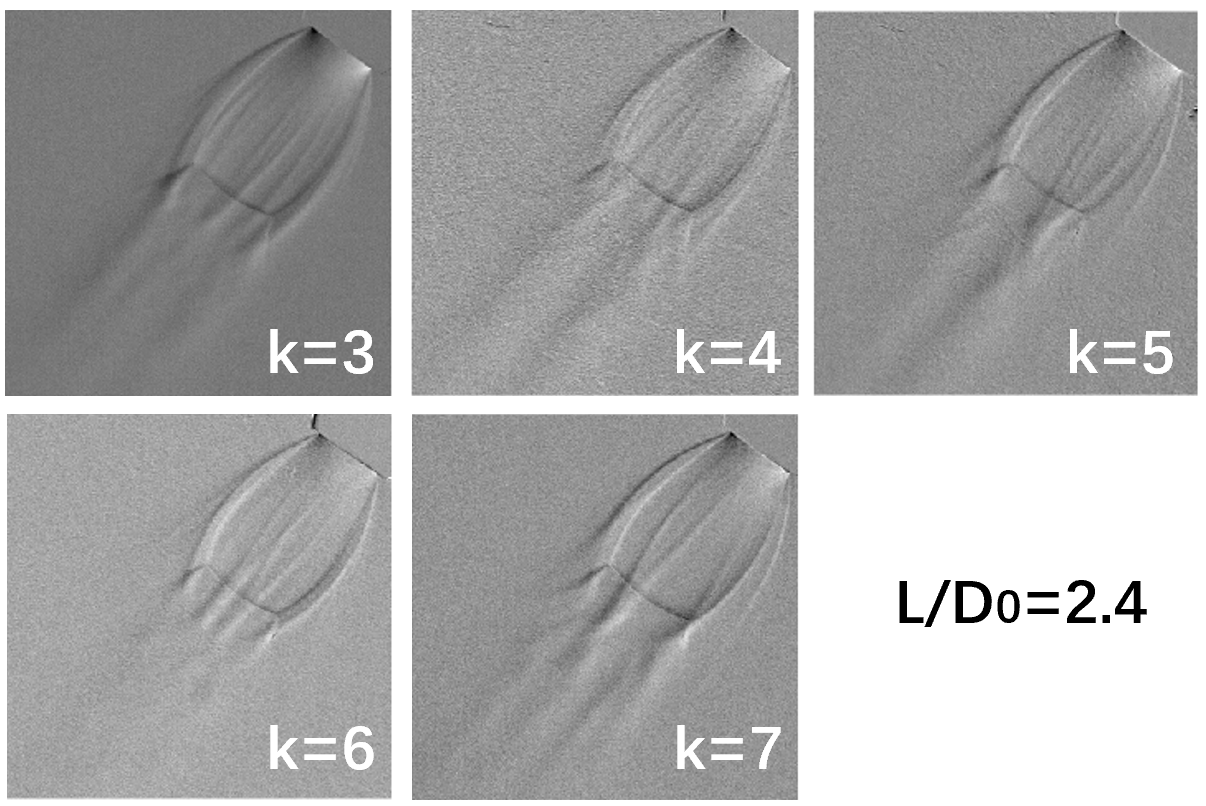}
\end{center}
\caption{\textbf{}Schlieren images of different activating wavenumber jets under unified conditions of L/D0=2.4.}
\label{fig:5}
\end{figure}

\begin{figure}[H]
\begin{center}
\includegraphics[width=\linewidth]{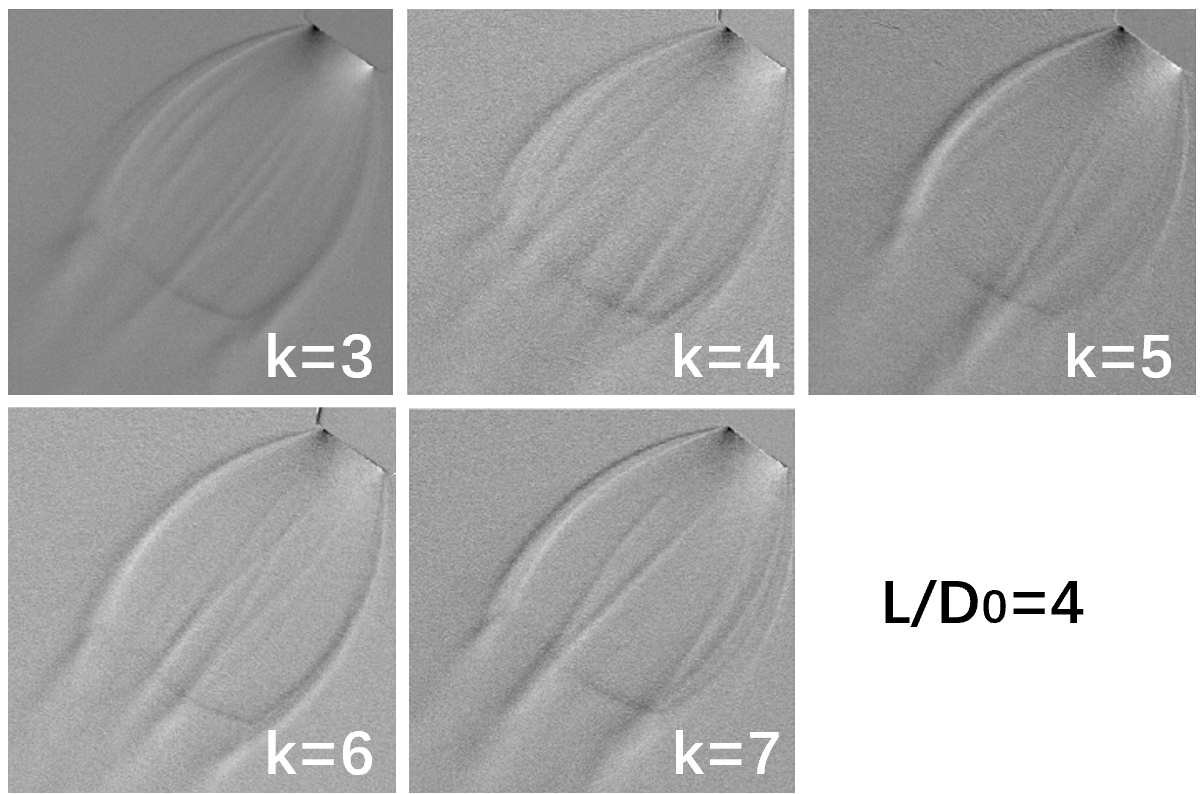}
\end{center}
\caption{\textbf{}Schlieren images of different activating wavenumber jets under unified conditions of L/D0=4}
\label{fig:6}
\end{figure}

Note that the number of streaks observed in jets activated by low-wavenumber disturbances is greater than the pre-designed excitation number. This suggests that low-wavenumber modes have an inherently low aerodynamic growth rate, which is insufficient to suppress the development of higher-order modes. In comparison, jets activated by high-wavenumber disturbances exhibit a relatively "clean" streak pattern. With appropriate geometric projection, the positions of the streaks in the images correspond to the locations of the disturbances at the nozzle exit, as shown in Fig. 7. For each peak in the disturbance, two homologous streaks are generated and later merge near the primary Mach disk.

\begin{figure}[H]
\begin{center}
\includegraphics[width=\linewidth]{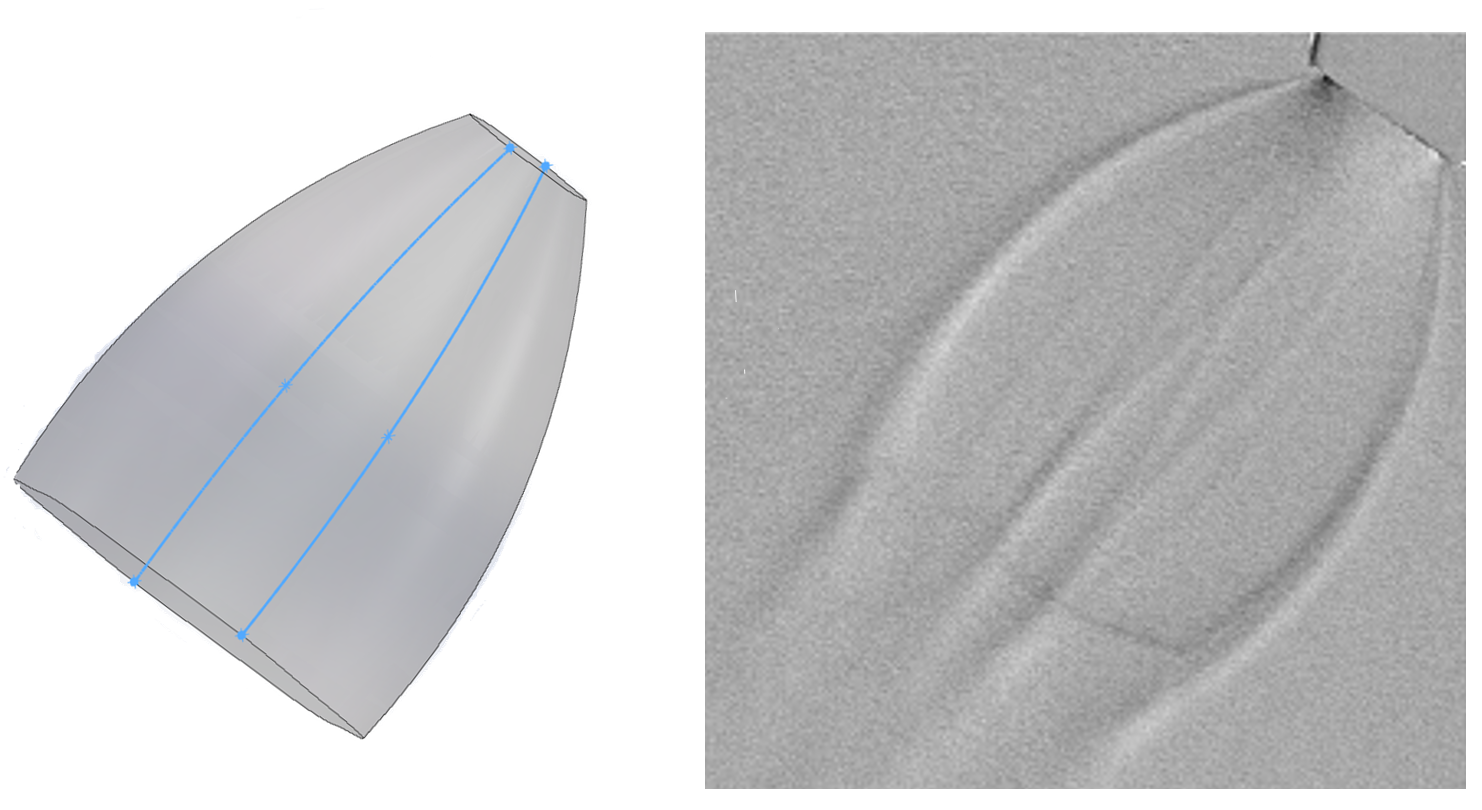}
\end{center}
\caption{\textbf{}Schematic diagram of the streamline at the peak of the disturbance, k=6}
\label{fig:7}
\end{figure}

As a path-integrated imaging method, schlieren did not provide spatially-resolved measurement of the detailed features of these streamwise streaks. To address this, we attempted to use high-temperature acetone vapor and a laser light sheet at a wavelength of 283 nm to observe a 45-degree slice of the flow field, aiming for clearer imaging of the shear layer. Unfortunately, although the boundaries of the shock cells were clearly visible in the PLIF images, the current signal strength was insufficient to extract detailed information from the jet shear layer. Further exploration on alternative PLIF or Mie scattering techniques is currently in progress, with an aim to achieve clearer visualization of the streamwise streaks in the shear layer, and to increase the activating wavenumber to better match the streak patterns observed in the smooth nozzle experiments.

\begin{figure}[H]
\begin{center}
\includegraphics[width=\linewidth]{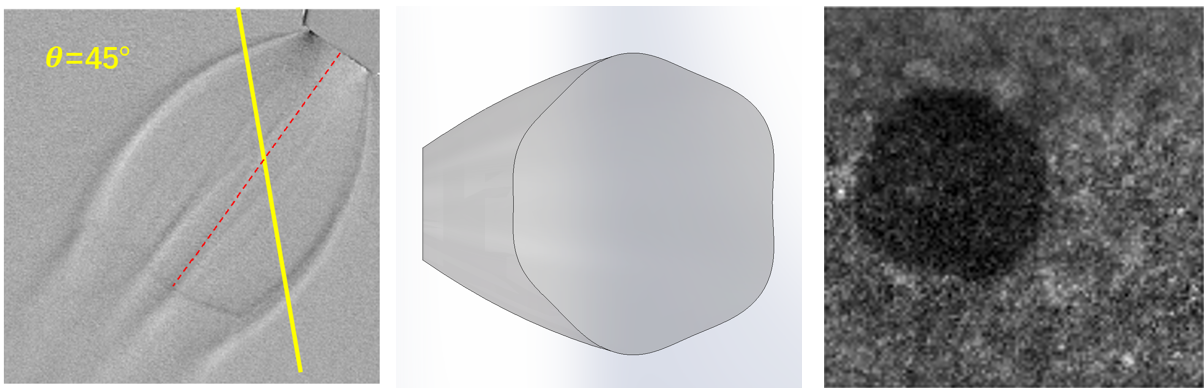}
\end{center}
\caption{\textbf{}Schematic diagram of the laser light sheet, and a representative PLIF image}
\label{fig:8}
\end{figure}

\section{Conclusions}

In summary, this study has demonstrated that the formation of streamwise streaks in underexpanded jets from round sonic nozzles is strongly influenced by geometric perturbations at the nozzle exit. Through systematic experiments using both smooth and intentionally perturbed nozzles, it was found that even minute surface roughness can give rise to stable streak patterns, as evident from the consistent rotation of these patterns with the nozzle orientation. Furthermore, the investigation into different modal perturbations revealed that higher-wavenumber disturbances more effectively shape the streamwise streak structures, whereas lower-wavenumber and smooth configurations are dominated by residual roughness effects. These findings highlight the critical role of nozzle roughness in the development of steamwise flow structures and provide insights for the design and operation of supersonic wind tunnels, where control over noise and flow patterns is essential.

\end{document}